\documentclass[%
superscriptaddress,
twocolumn,
amsmath,
amssymb,
aps,prl
]{revtex4-1}

\usepackage{graphicx}  
\usepackage{dcolumn}  
\usepackage{bm}            
\usepackage{amsmath}
\usepackage{xfrac}
\usepackage{xcolor}
\usepackage[colorlinks=true, citecolor=red]{hyperref}   
\usepackage{hhline}
\usepackage{diagbox}
\usepackage{siunitx}
\usepackage{enumitem}
\usepackage[normalem]{ulem}

\usepackage{ragged2e}

\begin{document}
	
	\title{Flavor symmetry breaking in spin-orbit coupled bilayer graphene}

	\author{Ming Xie}
	\affiliation{Condensed Matter Theory Center and Joint Quantum Institute, Department of Physics, 
			                University of Maryland, College Park, Maryland 20742, USA}
		                
	\author{Sankar Das Sarma}
    \affiliation{Condensed Matter Theory Center and Joint Quantum Institute, Department of Physics, 
    		                University of Maryland, College Park, Maryland 20742, USA}
    	                
	\date{\today}
	
	\begin{abstract}
		
		Recent experimental discovery of flavor symmetry breaking metallic phases in
		Bernal-stacked bilayer graphene points to the strongly interacting nature of electrons near the top (bottom) of its valence (conduction) band.
		Superconductivity was also observed in between these symmetry breaking phases when the graphene bilayer is placed under a small in-plane magnetic field or in close proximity to a monolayer WSe$_2$ substrate.
		Here we address the correlated nature of the band edge electrons and obtain the quantum phase diagram of their many-body ground states
		incorporating the effect of proximity induced spin-orbit coupling.
		We find that in addition to the spin/valley flavor polarized half and quarter metallic states, 
		two types of intervalley coherent phases emerge near the phase boundaries between the flavor polarized metals.
       Both spin-orbit coupling and in-plane magnetic field disfavor the spin-unpolarized valley coherent phase.
	  Our findings suggest possible competition between intervalley coherence and superconducting orders, arising from the intriguing correlation effects in bilayer graphene in the presence of spin-orbit coupling.
		
	\end{abstract}
	
	\maketitle

\emph{Introduction}.---\noindent
Near the Dirac points, electrons in graphene mono- and multi-layer structures 
acquire an emergent valley degree of freedom \cite{NetoGraphene, SankarRMP,BLG}.
The combined spin-valley flavor symmetry is often broken spontaneously
when the electron-electron Coulomb interaction is enhanced
relative to the kinetic energy of individual electrons.
One of the first experimental manifestations of such broken symmetries
 was the appearance of integer Hall conductance plateaus 
outside of graphene's normal four-fold degenerate quantum Hall sequences
\cite{AHMQuantumHallFM, KunYang, Kim2006, Eva2009, YoungGraphene, TutucBLG}.
Here electron kinetic energy is completely quenched by a strong magnetic field.
More recently,  the experimental discovery of a plethora of strongly correlated quantum phases
\cite{CaoInsulator, CaoSuper, Efetov, YoungDean, Gordon, YoungQAHE}
in twisted bilayer graphene unveiled the magic of moir\'{e} band engineering
that leads to a pair of nearly-flat bands when twist angle is close to 1$^{\circ}$.
The set of interaction-induced insulating ground states at integer moir\'{e} band fillings 
exemplifies the richness of the quantum phenomena that can arise from correlation-induced spontaneous flavor symmetry breaking.

Not long after the moir\'{e} paradigm, people observed flavor symmetry breaking metallic states, 
e.g., half- and quarter-metals,
in both (non-moir\'e) Bernal-stacked bilayer graphene (BLG) \cite{Young2022, Ashoori2022, Weitz2022, Stevan2023, Ludwig2023} and rhombohedral-stacked trilayer graphene \cite{YoungTLG},
indicating their strongly interacting nature at low electron or hole doping.
However, different from previous examples, these non-moir\'e graphene multilayers do not host narrow or flat bands.
The enhanced interaction is believed to originate from the finite bandgap induced by the vertical displacement field and the associated density of states including van Hove singularities \cite{Silva2008}.
More interestingly, proximate to the transition regions between different metallic phases, superconductivity was also reported
\cite{Young2022, Stevan2023, Ludwig2023, YoungTLGSC}.
In particular, in the bilayer case, the superconducting state was found to be
stabilized by either a small in-plane magnetic field \cite{Young2022} or by close proximity to a WSe$_2$ substrate \cite{Stevan2023, Ludwig2023}.
The nature of the flavor symmetry breaking phases and its relation to superconductivity \cite{ChouPhonon,ChouTunneling, NarangSC, LevitovSC, LevitovFM} remains unclear.

\begin{figure}[t!]
	\centering
	\includegraphics[width=0.38\textwidth]{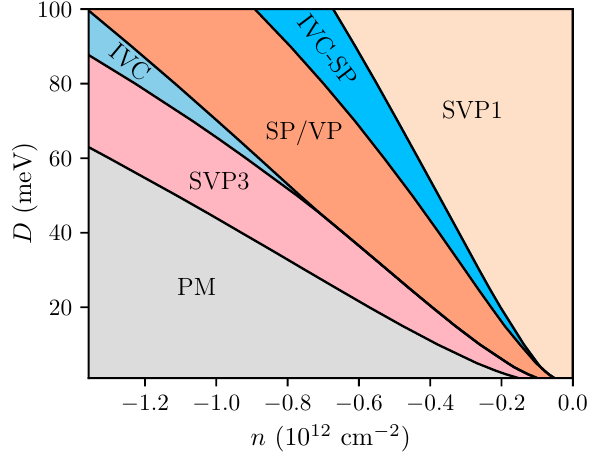}%
	\caption{Phase diagram of Bernal-stacked bilayer graphene as a function of 
		interlayer electric potential $D$ and carrier density $n$.
		We identify spin-valley polarized (SVP) phases with one (SVP1) or three (SVP3) partially filled flavors 
		and spin (or valley) polarized phases (SP/VP) with two partially filled flavors.
		The intervalley coherent (IVC) state is spin un-polarized while the IVC-SP state has a fixed spin polarization in each valley. The interaction parameters are taken to be $d=30$ nm and $\epsilon=10$.}
	
	\label{fig:phasediagram}
\end{figure}

In this Letter, we provide a full account of the quantum phase diagram of BLG under the influences of
screening by nearby metallic gates and spin-orbit coupling (SOC) induced by a WSe$_2$ substrate, following recent experiments \cite{Young2022, Ashoori2022, Weitz2022,Stevan2023}.
In addition to the spin/valley polarized half- and quarter-metals, we identify two types of
 intervalley coherent
phases near the transitions between the metallic states.
The SOC breaks the $SU(2)$ spin rotational symmetry in each valley and therefore reduces the
degeneracies of the flavor symmetry breaking ground states.
The intervalley coherent state without net spin-polarization is suppressed by the SOC,
pointing to possible competition with superconductivity which is enhanced by spin-orbit coupling \cite{Stevan2023}.

\emph{Symmetries and Mean-field Hamiltonian}.---\noindent
At low energies, the non-interacting electronic structure of BLG
per spin ($\uparrow$/$\downarrow$) and valley ($+$/$-$)
can be accurately described by a four-band effective model in the layer and sublattice basis.
Including the next-nearest neighbor and the third-nearest neighbor interlayer hopping terms splits
the quadratic band touching into four linear band crossings located around the Dirac point,
and induces a divergence, i.e., the van Hove singularity, in the density of states (DOS).
Application of an interlayer electric potential, $D$,
gaps out the band crossings and results in flattening of the dispersion near both bandedges.
This field-induced effective band-flattening is the mechanism for strong correlation effects in this non-moire system.
The parameters of the Hamiltonian are taken from Ref.~\cite{JeilTB} and detailed descriptions of the non-interacting bandstructure 
entering our theory is
given in the Supplemental Material \cite{SM}.

In the absence of SOC,
BLG has time reversal symmetry $T$, total charge conservation $U(1)_c$, 
valley charge conservation $U(1)_{\pm}$,
and independent spin rotational symmetry $SU(2)_{\pm}$ in each valley.
The interaction Hamiltonian takes the form
\begin{align}
	\mathcal{H}_I = \frac{1}{2A}\sum_{\alpha,\beta}\sum_{\bm{k},\bm{k}',\bm{q}} V(\bm{q}) 
	\hat{c}^{\dagger}_{\alpha,\bm{k}+\bm{q}}\hat{c}^{\dagger}_{\beta, \bm{k}'-\bm{q}}\hat{c}_{\beta,\bm{k}'}\hat{c}_{\alpha,\bm{k}},
	\label{intHam}
\end{align}
and respects all the symmetries of BLG. 
In Eq. (\ref{intHam})$, \alpha, \beta$ are composite indices $\{l,\sigma, s, \tau\}$ 
of layer ($l=1,2$), sublattice ($\sigma=A,B$), 
spin ($s=\uparrow/\downarrow$) and valley ($\tau=\pm$), respectively.
$A$ is the area of the system.
We consider the gate-screened (long-ranged) Coulomb interaction  
$V(\bm{q})=2\pi e^2/(\epsilon q) \tanh{(qd)}$
where $\epsilon$ is the dielectric constant of the environment 
and $d$ is the distance to the dual metallic gates.
We take both $\epsilon$ and $d$ as variables in our theory to control
the strength of Coulomb interaction.

The mean-field Hamiltonian includes the electrostatic Hartree contribution, 
accounting for the electric potential generated by the interlayer charge polarization,
and the Fock (exchange) contribution,
\begin{align}
	\mathcal{H}_{F}(\bm{k})=
	\sum_{i,j=0}^{3} \Delta_{ij}(\bm{k}) \tau_i\otimes s_j
\end{align}
where $\tau_i$ and $s_j$ are Pauli matrices for the valley and spin degrees of freedom.
$\Delta_{ij}(\bm{k})$ is a $4\times 4$ matrix whose elements are given by
$[\Delta_{ij}(\bm{k})]_{(l'\sigma'),(l\sigma)} =-\sum_{\bm{k}'} V(\bm{k}-\bm{k}')
\langle\hat{c}_{l,\sigma,s,\tau,\bm{k}'}^{\dagger}\tau_i^{\tau\tau'} s_j^{ss'}\hat{c}_{l',\sigma',s',\tau',\bm{k}'}\rangle/4A$,
where repeated indices are summed over.
We perform self-consistent mean-field calculations to study the evolution of 
the ground states as the carrier density and the interlayer potential are independently tuned.

\emph{Flavor Symmetry Breaking Phase Diagram}.---\noindent
Our theoretical approach is guided by experimental observations of
flavor symmetry breaking metallic states in the low density regime
where Coulomb interaction becomes important.
Fig.~\ref{fig:phasediagram} summarizes the phase diagram as a function of the
interlayer potential $D$ and the charge density $n$,
assuming typical experimental values for the gate distance $d=30$ nm and dielectric constant $\epsilon=10$.
In the high density limit, the ground state, as expected, is the normal Fermi liquid with a paramagnetic (PM) response
because of the dominance of the kinetic energy over Coulomb interaction.

As the hole density decreases, the ground state undergoes consecutive
correlation-driven transitions into 
a cascade of flavor ordered phases.
The ground states with spin polarization (SP) or valley polarization (VP) 
partially occupy two flavors, leaving the other two flavors empty of holes,
and are degenerate in energy. 
The SP state is a spin magnet with extra degeneracies due to the $SU(2)_{\pm}$ symmetry.
The VP state, in contrast, has a net orbital magnetization arising from the valley-contrasting Berry curvature.
On the other hand, the spin-valley polarized (SVP) ground states spontaneously break both spin and valley degeneracies
and partially occupy either one (SVP1) or three (SVP3) out of the four flavors.
They have both spin and orbital magnetizations and are expected to exhibit anomalous Hall effect due to nonzero Berry curvatures.

In the transition regions between the SVP and the SP/VP states,
two types of intervalley coherent states emerge.
The IVC state is spin-unpolarized
and typically has two identical (valley coherent) Fermi surfaces for the two spin flavors.
The IVC-SP state inherits the tendency toward spin polarization at low densities 
and has only one Fermi surface.
Both the IVC and IVC-SP states have degeneracies generated by the $SU(2)_+\times SU(2)_-$ spin rotation.

\begin{figure}[t!]
	\centering
	\includegraphics[width=0.48\textwidth]{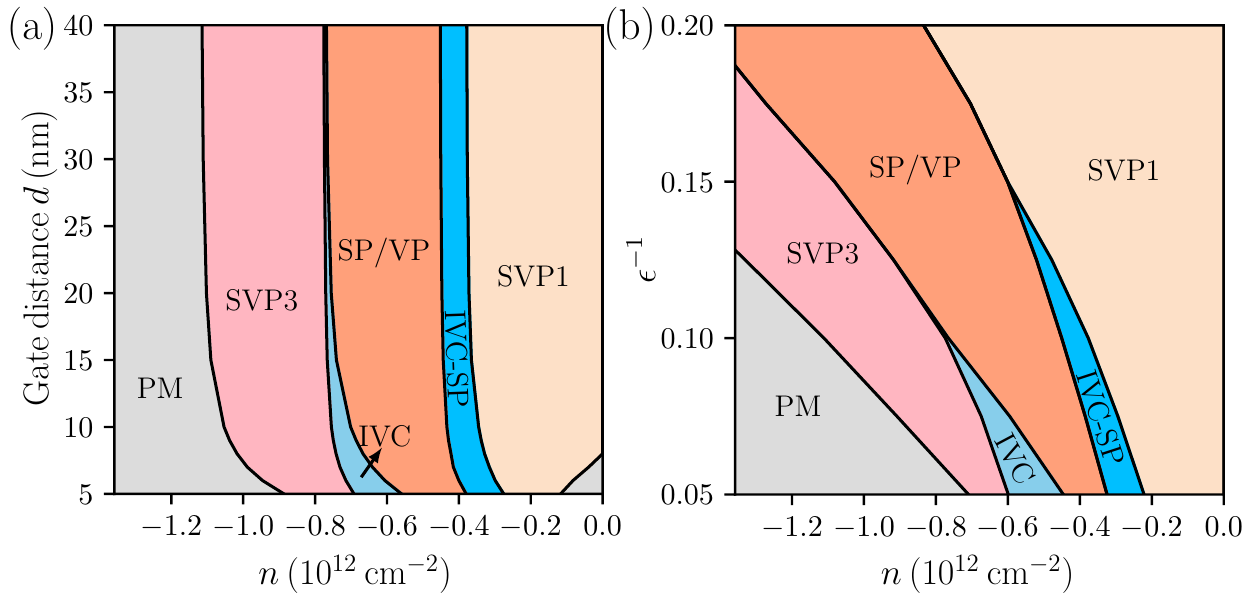}%
	\caption{Dependence of the phase diagram on Coulomb interaction parameters: 
		(a) screening by metallic gates at a distance $d$ and 
		(b) environmental dielectric constant $\epsilon$.
		The color scheme is the same as in Fig. \ref{fig:phasediagram}.
		In both (a) and (b), the interlayer potential is set at $D=50$ meV;  
		$\epsilon=10$ is used in (a) and $d=30$ nm is used in (b).
	}
	
	\label{fig:vseps}
\end{figure}

\emph{Screening}.---\noindent
The Coulomb interaction strength in our theory is controlled by
the distance $d$ to the dual metallic gates
and the dielectric constant $\epsilon$ of the barrier material.
The gate screening suppresses the long range part of the Coulomb interaction with momentum $q<d^{-1}$.
As $d$ increases, the flavor symmetry breaking transitions shift toward higher hole densities as shown in Fig. \ref{fig:vseps}(a).
When $d$ reaches a value about $d_c=15$ nm, the phase diagram becomes nearly independent of $d$,
suggesting the critical density ($D$ dependent) for the cascade of transitions is $d_c^{-2}=0.4\times 10^{12}$ cm$^{-2}$.
When the inter-electron spacing becomes less than $d_c$, the gate screening no longer affects the phase transitions.
We note that the paramagnetic Sym-12 state observed in experiments \cite{Young2022, Ludwig2023, Stevan2023} appears in the phase diagram 
at small gate distances (Fig.~\ref{fig:vseps}(a)), suggesting that the Sym-12 state signifies strongly screened Coulomb interaction.

Unlike the gate screening, 
the dielectric constant suppresses the full range of the Coulomb interaction 
uniformly independent of inter-electron spacing. 
As a consequence, when $\epsilon^{-1}$ (proportional to the interaction strength) increases, 
the flavor symmetry breaking phases expand to ever larger hole densities as shown in Fig. \ref{fig:vseps}(b).
Most importantly, Fig. \ref{fig:vseps}(b) shows that increasing the interaction strength disfavors
the intervalley coherent phases, including both the IVC and the IVC-SP phases, relative to the flavor polarized phases.
Compared to flavor polarization, 
 forming intervalley coherence saves kinetic energy 
at the cost of gaining less exchange interaction energy; therefore,
increasing interaction strength will eventually favor flavor polarized phases.

\begin{figure}[t!]
	\centering
	\includegraphics[width=0.499\textwidth]{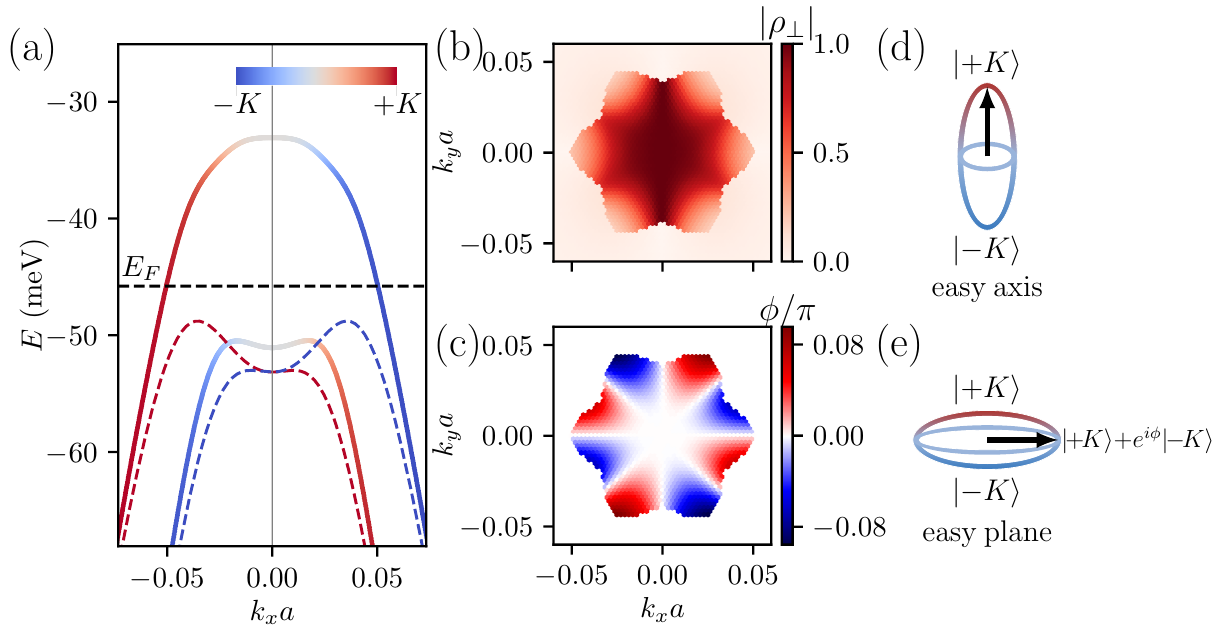}%
	\caption{(a) Self-consistent mean-field bandstructure of the IVC-SP ground state at
					      $n=-0.27\times 10^{12}$ cm$^{-2}$ and $D=30$ meV. 
		                  Solid (dashed) line represents the dispersion of the occupied (empty) spin flavor. 
		                  and the line color indicates the wavefunction weight on each valley.
		                  The black dashed line marks the Fermi energy.
					(b-c) The amplitude (b) and phase (c) of the intervalley coherence order parameter 
					     for the IVC-SP ground state shown in (a).
					(d-e) Schematic illustrations of the two types of valley anisotropy scenario: 
					    (d) easy axis and (e) easy plane.
				}
	
	\label{fig:IVC}
\end{figure}

\emph{Intervalley Coherence}.---\noindent
In the transition regions between the SVP phases and the SP/VP phases,
the ground state tends to form spontaneous coherence between the two valleys,
giving rise to valley hybridized bands.
Fig. \ref{fig:IVC}(a) plots the typical bandstructure for the IVC-SP phase 
where only one spin flavor is occupied (solid line).
For the un-occupied spin flavor (dashed line), no intervalley coherence is established
and the bands from the two valleys remain decoupled.
On the other hand, for the IVC phase, the intervalley coherence is equal in magnitude 
for both spin copies.
For both the IVC-SP and the IVC phases,
because of the $SU(2)_+\times SU(2)_-$ symmetry,
independent rotation of spin directions in each valley generates degenerate ground states,
and the spin polarization directions in the IVC-SP phase, for example, can be different in the two coupled valleys.

\begin{figure}[t!]
	\centering
	\includegraphics[width=0.5\textwidth]{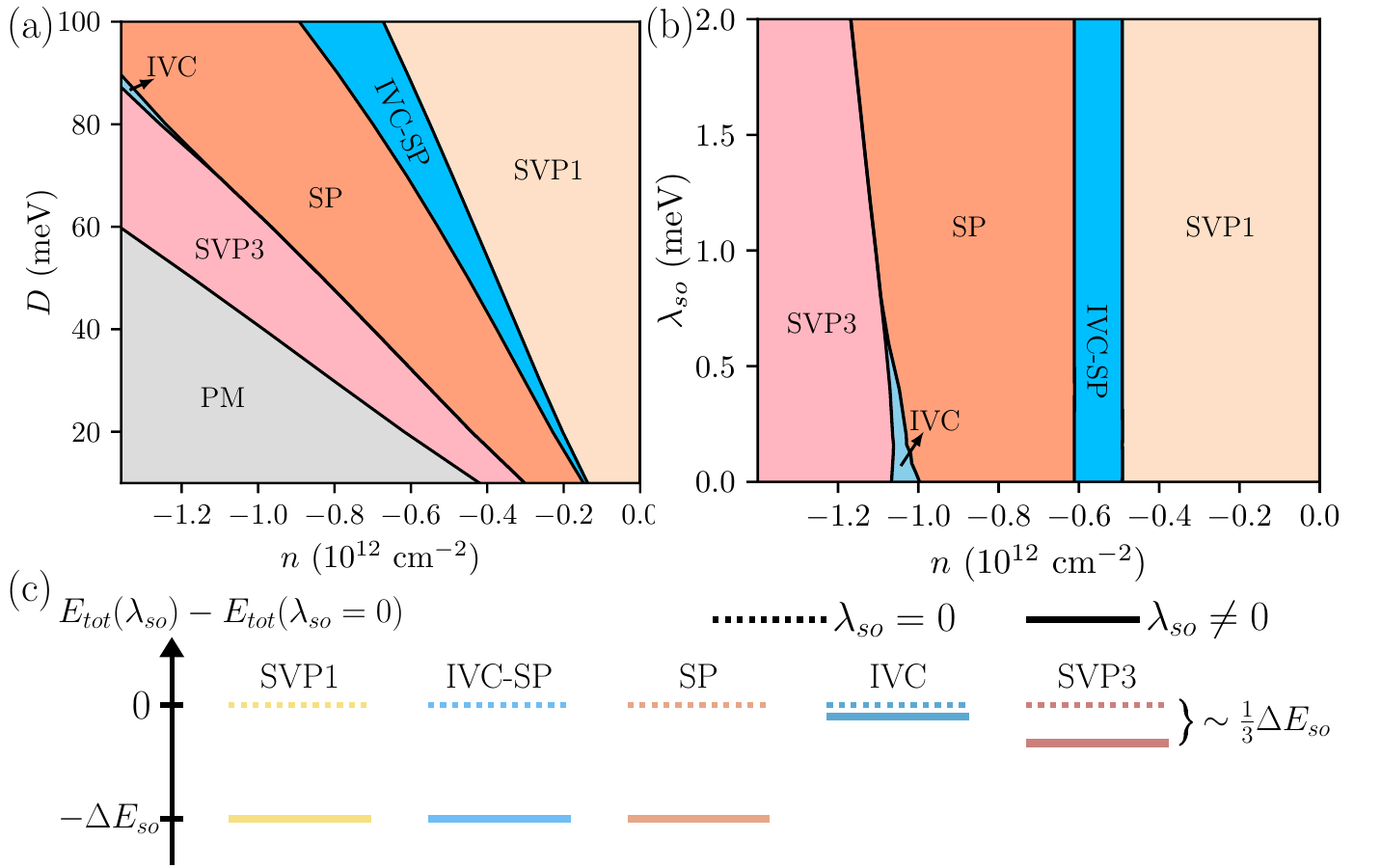}%
	\caption{(a) Phase diagram of bilayer graphene with proximity induced Ising SOC.  
		The SOC strength is $\lambda_{so}=1.0$ meV. 
		The color scheme is the same as in Fig. \ref{fig:phasediagram}.
		(b) Dependence of the phase diagram on the SOC strength $\lambda_{so}$.
		The interlayer potential is fixed at a typical value of $D=70$ meV.
		(c) Schematic illustration of the total energy differences without and with the presence of the Ising SOC.
		$\Delta E_{so} \approx |n|\lambda_{so}/2$ for the SVP1, IVC-SP, and SP phases.
	}
	
	\label{fig:ising}
\end{figure}

To understand the symmetry of the intervalley coherence, 
we define the order parameter:
$\rho_{ss'}^{\rm IV}(\bm{k})=\langle\hat{c}_{+,s',\bm{k}}^{\dagger} \hat{c}_{-,s,\bm{k}}\rangle$
where $\hat{c}_{\tau,s,\bm{k}}^{\dagger}$ is a vector in sublattice and layer basis.
We can decompose $\rho^{\rm IV}$ as:
\begin{align}
	\rho^{\rm IV}(\bm{k}) = \rho_{\perp}(\bm{k}) \Pi
\end{align}
where $\rho_{\perp}(\bm{k})$ is the valley coherence and 
$\Pi$ describes the spin structure in the two valleys.
Fig. \ref{fig:IVC}(b) and (c) provide the amplitude and phase of $\rho_{\perp}(\bm{k})=|\rho_{\perp}(\bm{k})|e^{i\phi_{\bm{k}}}$ as a function of momentum.
The amplitude $|\rho_{\perp}(\bm{k})|$ drops sharply to nearly zero at the Fermi surface (Fig. \ref{fig:IVC}(b)).
The phase $\phi_{\bm{k}}$ varies only slightly around a spontaneous value $\phi_{\bm{k}}\approx\phi$
and the ground state is degenerate for arbitrary angle $\phi$.

To best illustrate the $SU(2)_{\pm}$ symmetry,
we express $\Pi$ in terms of the spin $SU(2)$ rotation operator:
$\mathcal{R}_{\pm}=e^{i\theta_{\pm} \bm{n}_{\pm}\cdot\bm{s}/2}$,
acting on the $+K$ and $-K$ valleys, respectively.
$\bm{n}_{\pm}$ are arbitrary unit vectors.
For the IVC-SP phase, we have only one copy of spin polarization occupied,
$\Pi=\mathcal{R}_+|+\negthickspace\uparrow\rangle\langle-\negmedspace\uparrow\negthickspace|\mathcal{R}^{-1}_-$.
For the IVC phase, the two orthogonal spin copies are equally occupied,
$\Pi=\mathcal{R}_+|+\negthickspace\uparrow\rangle\langle-\negmedspace\uparrow\negthickspace|\mathcal{R}^{-1}_-
       +\mathcal{R}_+|+\negthickspace\downarrow\rangle\langle-\negmedspace\downarrow\negthickspace|\mathcal{R}^{-1}_-$.
The ground states generated using arbitrary  $\bm{n}_{\pm}$ and $\theta_{\pm}$ are all degenerate in the absence of SOC.
Intervalley coherent states have also been identified theoretically in rhombohedral 
trilayer graphene \cite{ChatterjeeTLG, chunliTLG, Ashvin}, 
where the order parameter has a non-zero phase winding \cite{ChatterjeeTLG} 
in contrast to the nearly uniform phase here.

%

The competition between flavor polarized phases and intervalley coherent phases
 can be understood in terms of  $n$ and $D$ dependent valley anisotropy.
Fig. \ref{fig:IVC}(d) and (e) schematically illustrate two types of valley anisotropy scenarios.
Valley polarized phases are favored by easy axis valley anisotropy. 
Intervalley coherent phases, corresponding to in-plane valley ordering, 
are favored when valley anisotropy is easy plane type, 
analogous to an XY magnet.
One can write down a phenomenological Ginzburg-Landau energy functional as
\begin{align}
	\mathcal{E}_{\rm ani}=K_{\perp} |\rho_{\perp}|^2 + K_z\rho_z^2
\end{align}
where $\rho_z=\rho_+-\rho_-$ is the valley polarization density. $K_{\perp}$ and $K_z$ are in-plane and out-of-plane valley anisotropy
constants.
Based on mean-field calculations, we find that the ground state energy is isotropic with respect to in-plane polarization angle $\phi$.
Similar to spin polarization in a ferromagnet, 
the anisotropy constant toward valley-polarization takes 
the form $K_z=\mathcal{N}^{-1}_0-I_{ex}^{VP}$ 
where $\mathcal{N}_0$ is the DOS per flavor at Fermi energy
and
$I_{ex}^{VP} \propto V\langle\hat{c}_{\tau }^{\dagger} \hat{c}_{\tau}\rangle$
is the exchange interaction constant.
$K_z<0$ corresponds to the Stoner criteria for valley-polarization instability.
On the other hand, forming valley coherence does not cost kinetic energy
but instead gains an exchange energy so that
$K_{\perp}=-I^{IV}_{ex}$,
where $I^{IV}_{ex} \propto V \langle\hat{c}_{\tau}^{\dagger} \hat{c}_{-\tau}\rangle$.
When $K_{\perp}<K_{z}$ and $K_{\perp}<0$,
the ground state prefers to have finite $\rho_{\perp}$;
when $K_{z}<K_{\perp}$ and $K_z<0$,
valley polarization is preferred.
The competition between the kinetic energy cost and the exchange energy gain determines the type of valley anisotropy,
and therefore selects the ground state
as $D$, $n$ and interaction strength are varied.




\emph{Ising spin-orbit coupling}.---\noindent
Placing the graphene bilayer in close proximity to a WSe$_2$ substrate
introduces SOC via interfacial van der Waals interaction.
The induced SOC includes two contributions:
the Ising contribution inherited from the WSe$_2$ layer and the Rashba contribution arising from inversion symmetry breaking at the interface.
Because the Rashba SOC is off-diagonal in sublattice,
it has negligible effect on the states near the band-edges of bilayer graphene where electronic wavefunction is sublattice polarized.
Therefore, we focus on the Ising SOC which is diagonal in sublattice and has dominant effect on the band-edge states.

The Ising SOC Hamiltonian is given by
\begin{align}
	H_{so} = \frac{1}{2}\lambda_{so}\tau_z s_z P_{l=1}
\end{align}
where $\lambda_{so}$ is the SOC strength and
$P_{l=1}$ projects to the bottom layer ($l=1$) which is closest to the WSe$_2$
substrate.
$H_{so} $ introduces a spin quantization axis
perpendicular to the layer and breaks the $SU(2)_{\pm}$ symmetries.

Fig.~\ref{fig:ising}(a) plots the mean-field phase diagram in the presence of the Ising SOC.
Different from the diagram in Fig.~\ref{fig:phasediagram} in the absence of Ising SOC, 
the ground states no longer have spin rotational symmetries.
The Ising SOC selects a particular spin-valley configuration for each type of
flavor symmetry breaking phases as listed in Table~\ref{tab:flavor}.
Because of the flavor-dependent band splitting caused by Ising SOC, the three occupied flavors in the SVP3 phase have un-equal hole populations
$(n/3+\delta n, n/3+\delta n, n/3-2\delta n)$.
For the IVC phase, the Ising SOC tilts the valley order to have a small out-of-plane component, $\pm \delta n$,
which is opposite for opposite spins \cite{IVCnote}.
In both cases, $\delta n \propto \mathcal{N}_0 \lambda_{so}$ when $\lambda_{so}$ is small (See Fig. S4 of the Supplemental Material).

The most prominent feature in Fig.~\ref{fig:ising}(a) is the suppression of the IVC phase
when Ising SOC is included.
To best illustrate this trend, we plot the phase diagram as a function of $\lambda_{so}$ in Fig.~\ref{fig:ising}(b).
It can be understood approximately assuming that the Ising SOC splitting projected to the band basis 
is moment independent, which is indeed the case for states near the band-edges \cite{SM}.
The total energy decrease caused by $H_{so}$ is
nearly identical, $\Delta E_{so}\approx |n|\lambda_{so}/2$, for the SVP1, SP and IVC-SP phases as 
schematically shown in Fig.~\ref{fig:ising}(c).
In the IVC phase, the total energy change is dominated by kinetic energy cost associated with the the out-of-plane tilting
and is a second order correction $\propto \mathcal{N}_0\lambda_{so}^2$
(See Fig. S4 of the Supplemental Material).
For the SVP3 phase, the total energy change to the first order in $\lambda_{so}$ is roughly $-\Delta E_{so}/3$.
Clearly the IVC phase is disfavored relative to its neighboring phases as $\lambda_{so}$ increases.
The SVP1/IVC-SP and the IVC-SP/SP phase boundaries, however, remain nearly unchanged because these phases 
gain the same amount of energy from Ising SOC.

\begin{table}[t!]
	\begin{tabular}{ >{\centering\arraybackslash}p{1.1cm}  >{\centering\arraybackslash}p{3.6cm}  >{\centering\arraybackslash}p{3.5cm}}
		\hhline{===}
		Phases\vspace{0.5em} & $\lambda_{so}=0$ &  $\lambda_{so} \neq 0$ \tabularnewline \hline
		SVP1     & $\mathcal{R}_+|+\negthickspace\uparrow\rangle \langle+\negthickspace\uparrow\!|\mathcal{R}^{-1}_+$
		&$|+\negthickspace\downarrow\rangle \langle+\negthickspace\downarrow\!|$ \\ 
		IVC-SP& $\mathcal{R}_+|+\negthickspace\uparrow\rangle \langle-\negthickspace\uparrow\!|\mathcal{R}^{-1}_-$
		&$|+\negthickspace\downarrow\rangle \langle-\negthickspace\uparrow\!|$ \\ 
		SP   & $\sum_{\tau}\mathcal{R}_{\tau}|\tau\negthickspace\uparrow\rangle \langle\tau\negthickspace\uparrow\!|\mathcal{R}^{-1}_{\tau}/2$
		&$(|+\negthickspace\downarrow\rangle \langle+\negthickspace\downarrow\!|+
		|-\negthickspace\uparrow\rangle \langle-\negthickspace\uparrow\!|)/2$ \\ 
		IVC     & $\sum_{s}\mathcal{R}_{+}|+\negthickspace,s\rangle \langle -\negthickspace,s|\mathcal{R}^{-1}_{-}/2$
		&$(|+\negthickspace\uparrow\rangle \langle-\negthickspace\uparrow\!|-
		|+\negthickspace\downarrow\rangle \langle-\negthickspace\downarrow\!|)\eta$ \\ 
		SVP3     & $(I-\mathcal{R}_+|+\negthickspace\uparrow\rangle \langle+\negthickspace\uparrow\!|\mathcal{R}^{-1}_+)/3$
		&$(I-|+\negthickspace\uparrow\rangle \langle+\negthickspace\uparrow\!|)/3$ \\ \hhline{===}
	\end{tabular}
	\caption{Flavor configurations for different symmetry breaking phases without ($\lambda_{so}=0$) and with ($\lambda_{so} \neq 0$) proximity induced spin-orbit coupling. For flavor polarized phases, the table element represents the valley diagonal component of the density matrix; for intervalley coherent phases
	it represents valley off-diagonal component of the density matrix.
 	$\eta=\sin\theta\cos\theta $ represents the out-of-plane tilting of the valley pseudo-spin by an angle of $\theta$.
	(See Supplemental Material for detailed discussions.)
	$\mathcal{R}_{\pm}$ indicates spin rotation degeneracy of the ground states.}
		\label{tab:flavor}
\end{table} 

\emph{In-plane magnetic field}.---\noindent
Applying a small in-plane magnetic field has been shown to stabilize superconductivity \cite{Young2022}.
Here we consider its effect on flavor symmetry breaking.
Because of the small interlayer spacing in bilayer graphene,
a low in-plane magnetic field  has negligible orbital effect \cite{FalkoInplane, inplanefield} (also see Supplemental Material \cite{SM} for details).
What remains dominant is the Zeeman effect with magnitude $\sim 10^{-1}$ meV for $B\sim1T$.
The Zeeman Hamiltonian is
\begin{align}
H_{Z} = \frac{1}{2}E_Z s_z\tau_0	
\end{align}
where $E_Z$ is the Zeeman splitting.
Analogous to the effect of the Ising SOC,
$H_Z$ breaks the spin degeneracy of the ground state.
It selects a set of ground states with different spin configurations compared to those in Table~\ref{tab:flavor}.
However, the total energies for the flavor breaking quantum phases change in the same manner as shown in Fig.~\ref{fig:ising}(c) (see Supplemental Material \cite{SM} for detailed discussion).
Therefore we expect the phase diagram in the presence of such in-plane magnetic field will
look qualitatively the same as in Fig.~\ref{fig:ising} (a) and (b).
Note that the preferred spin axis under the Ising SOC and the in-plane field are orthogonal to each other.

\emph{Discussion}.---\noindent
Superconductivity in BLG so far
relies on either the proximity induced SOC or an applied small
in-plane magnetic field.
Our mean-field calculations find that both of these effects
suppress the IVC phase relative to its neighboring flavor polarized phases,
which points to possible competition between the IVC order and the superconducting order.
We believe that this suppression of the competing IVC phase is what leads to superconductivity in BLG.
Moreover, experimental observations of flavor symmetry breaking phases 
vary from device to device depending sensitively on the parameter details as shown in the current work. 
Our predicted phase diagrams should be tested in future experiments.

Flavor symmetry breaking transitions have also been found in magic angle bilayer graphene, which features a similar hierarchy of sequentially flavor broken phases. 
The superlattice gap limited one-electron-per-moiré-cell electron filling allows correlated insulating states separating the cascade of flavor transitions.

\begin{acknowledgments}
	{\em Acknowledgment.}---\noindent	The authors acknowledge helpful interactions with L. Holleis, A. F. Young, and Y.-Z. Chou. This work is supported by Laboratory for Physical Sciences.
\end{acknowledgments}


\begin{thebibliography}{99}
\bibitem{IVCnote}
	We find another intervalley coherent phase arising under finite Ising SOC and is close in energy with the IVC phase.
	In contrast to the IVC state, this phase has unequal populations in the two spin flavors. Its flavor configuration (defined in Table \ref{tab:flavor})
	is $[(1-\eta)|+\negthickspace\uparrow\rangle \langle-\negthickspace\downarrow\!|+
	(1+\eta)|+\negthickspace\downarrow\rangle \langle-\negthickspace\uparrow\!|]/2$, where $\eta\propto \mathcal{N}_o\lambda_{so}$ captures the unequal population between the two spin copies.
\end{thebibliography}
\end{document}